\documentclass[iop]{emulateapj}
\usepackage{amsmath}
\usepackage{rotating}

\def\stacksymbols #1#2#3#4{\def\theguybelow{#2}
        \def\verticalposition{\lower#3pt}
        \def\spacingwithinsymbol{\baselineskip0pt\lineskip#4pt}
        \mathrel{\mathpalette\intermediary#1}}
\def\intermediary #1#2{\verticalposition\vbox{\spacingwithinsymbol
        \everycr={}\tabskip0pt
        \halign{$\mathsurround0pt#1\hfil##\hfil$\crcr#2\crcr
                \theguybelow\crcr}}}

\shorttitle{Star Formation Quenching and Intermittent Cooling Flows}
\shortauthors{GUO}

\begin{document}
\bibliographystyle{apj} 

\title {Connecting Star Formation Quenching with Galaxy Structure and Supermassive Black Holes through Gravitational Heating of Cooling Flows}

\author{Fulai Guo\altaffilmark{1,2}}

\altaffiltext{1}{ETH Z\"{u}rich, Institute for Astronomy, Wolfgang-Pauli-Strasse 27, CH-8093, Z\"{u}rich, Switzerland; fulai.guo@phys.ethz.ch}

\altaffiltext{2}{Zwicky Prize Fellow}

\begin{abstract}

Recent observations suggested that star formation quenching in galaxies is related to galaxy structure. Here we propose a new mechanism to explain the physical origin of this correlation. We assume that while quiescent galaxies are maintained quenched by a feedback mechanism, cooling flows in the hot halo gas can still develop intermittently. We study cooling flows in a large suite of around 90 hydrodynamic simulations of an isolated galaxy group, and find that the flow development depends significantly on the gravitational potential well in the central galaxy. If the galaxy's gravity is not strong enough, cooling flows result in a central cooling catastrophe, supplying cold gas and feeding star formation to galactic bulges. When the bulge grows prominent enough, compressional heating starts to offset radiative cooling and maintains cooling flows in a long-term hot mode without producing cooling catastrophe. Our model thus describes a self-limited growth channel for galaxy bulges, and naturally explains the connection between quenching and bulge prominence. In particular, we explicitly demonstrate that $M_{*}/R_{\rm eff}^{1.5}$ is a good structural predictor of quenching. We further find that the gravity from the central supermassive black hole also affects the bimodal fate of cooling flows, and predict a more general quenching predictor to be $M_{\rm bh}^{1.6}M_{*}/R_{\rm eff}^{1.5}$, which may be tested in future observational studies. 
 
\end{abstract}


\section{Introduction}
\label{section:intro}

Large galaxy surveys have demonstrated that galaxies can be generally classified into two distinct populations: blue, star-forming or red, quiescent galaxies (e.g., \citealt{strateva01}; \citealt{kauffmann03a}). It is usually thought that quiescent galaxies evolve from star-forming galaxies through a process denoted as ``quenching". Many quenching mechanisms have been proposed in literature, including halo quenching (e.g., \citealt{birnboim03}; \citealt{keres05}), ``quasar-mode" active galactic nucleus (AGN) feedback (e,g., \citealt{dimatteo05}), ``radio-mode" AGN feedback (e.g., \citealt{croton06}), morphological quenching \citep{martig09}, ram pressure stripping \citep{gg72}, etc.

To explore the quenching mystery, many observational studies have been performed to link star formation (SF) activity with galaxy properties. Evidences have been accumulating that SF quenching is correlated with galaxy structure represented by various structural parameters including the effective stellar surface density $M_{*}/R_{\rm eff}^{2}$ (e.g., \citealt{kauffmann03} and \citealt{kauffmann06}), $M_{*}/R_{\rm eff}$ (e.g., \citealt{franx08}), the Sersic index $n$ (e.g., \citealt{bell08}), or central stellar surface density (\citealt{cheung12}; \citealt{fang13}). Here $M_{*}$ and $R_{\rm eff}$ are respectively the total stellar mass and effective half-light radius. More recently, \citet{omand14} show that the best predictor of quenching in central galaxies may be $M_{*}/R_{\rm eff}^{1.5}$. These studies suggest that quenching is related to the prominence of bulges. 

Observational correlations do not directly tell if these structural parameters are causative or merely correlative to quenching, and it is not yet fully clear what causes these correlations. Prominent bulges may result from major mergers, which trigger bulge-building starbursts and powerful AGN activity, quenching SF. Prominent bulges may also be necessary to quench SF in gas disks \citep{martig09}. In this letter, we propose and numerically investigate a new mechanism to connect SF quenching with prominent bulges through gravitational heating of cooling flows, which may develop intermittently in the hot gaseous halo.

\section{The Model and Numerical Methods}
\label{section2}

\subsection{The Model}

Cold gas is the fuel of SF in galaxies. We consider galaxies (including centrals and satellites) where the supply of cosmic cold gas has already been shut off by cosmic shock heating. Galaxies also contain hot gas, which could come from accretion of shock-heated cosmic gas or stellar outflows (e.g., \citealt{mathews03}). The hot gas cools by radiating thermal bremsstrahlung and metal line emissions, potentially serving as an internal channel of feeding cold gas to galaxies through cooling flows (\citealt{fabian94}; \citealt{peterson06}). Recent observations have ruled out unimpeded, long-lasting cooling flows in galaxy clusters \citep{peterson06}. Radio-mode AGN feedback is usually invoked to heat the hot gas (\citealt{mcnamara07}), serving as the ``maintenance" mode to keep quiescent galaxies quenched (\citealt{croton06}; \citealt{bower06}).

However, it is not guaranteed that radio-mode AGN feedback always continuously shuts off cooling flows. Observations of radio lobes and X-ray cavities in galaxy groups and clusters indicate that AGN events have duty cycles \citep{mcnamara07}, implying that AGN heating and the cooling-flow development are intermittent \citep{peterson06}. A reasonable scenario is that the development of cooling flows feeds the central supermassive black hole (SMBH) and triggers AGN activity, which suppresses cooling flows in return. The diminishing cooling flows then stop feeding the SMBH and turn off AGN activity, which restarts cooling flows, leading to a new feedback loop. Within this scenario, intermittent cooling flows operate in galaxies, and if resulting in cooling catastrophe, would deposit cold gas to galaxies, fueling SF as observed in many central galaxies in cool core clusters (\citealt{peterson06}; \citealt{odea08}).

Recent simulations in \citet{guo14} show that the cooling-flow development is bimodal in central regions, depending on the competition between cooling and compressional heating. Some systems evolve into a central cooling catastrophe quickly at about the central cooling time (often $<0.1$ Gyr), while other systems maintain the hot-mode flow with cuspy central temperature profiles for several Gyrs or permanently. An important factor determining the cooling flow fate is the gravitational potential well within the central few tens of kpc regions, which is often dominated by the central galaxy, and for the innermost regions, by the central SMBH.

While the long-term hot-mode flow may not trigger SF, the cold-mode cooling catastrophe supplies cold gas directly to galaxy bulges, resulting in SF and more prominent (compact) bulges. If the bulge of a newly quenched galaxy is not significant enough, cooling flows may result in short epochs of central cooling catastrophe separated by episodic AGN feedback events, leading to more prominent bulges until the central gravitational potential well is deep enough to maintain a long-term hot-mode cooling flow. This mechanism naturally results in a connection between galaxy quenching and bulge prominence represented by various structural parameters, e.g., $M_{*}/R_{\rm eff}^{2}$, $M_{*}/R_{\rm eff}^{1.5}$, the Sersic index, or central stellar mass density. In this paper, we explore such a ``theoretical" correlation with hydrodynamic simulations, and compare it with observations.

\subsection{Assumptions and Numerical Setup}

We performed a large suite of more than 90 hydrodynamic simulations to investigate how the fate of cooling flows depends on $M_{*}$, $R_{\rm eff}$, and the mass of the central SMBH, $M_{\rm BH}$. We consider the development of pure cooling flows in a spherically symmetric system -- galaxy group NGC 4261. The density and temperature distributions of hot gas in NGC 4261 have been determined by {\it Chandra} observations, and the dark matter halo and stellar distribution are also well constrained \citep{humphrey06}. The system setup and numerical methods are similar to those described in detail in \citet{guo14}. 

For radiative cooling, we use the \citet{sd93} cooling function, which depends on gas metallicity. In galaxy groups, gas metallicity is often near solar in central regions, and decreases to $Z\sim0.1Z_{\sun}$ within $r\sim r_{500}$ (\citealt{rp07}). For simplicity, we assume a spatially uniform gas metallicity $Z=0.8Z_{\sun}$. We turn off radiative cooling below a minimum temperature of $10^{4}$ K. 

The cooling-flow development is strongly shaped by gravitational heating in the  gravitational potential well, contributed by three static components: the dark matter halo, the central galaxy and SMBH. For the dark matter halo, we adopt a Navarro-Frenk-White profile \citep{navarro97} with the virial mass $M_{\rm vir}=6.7\times 10^{13} M_{\sun}$ and the scale radius $r_{\rm s}=281.1$ kpc \citep{humphrey06}. We assume the galaxy's mass distribution to have a Hernquist profile \citep{hernquist90} with the total stellar mass $M_ {*}$ and the half-light radius $R_{\rm eff}$. The SMBH's gravitational potential is assumed to be
\begin{eqnarray}
\Phi_{\rm BH}= -\frac{GM_{\rm BH}}{r-r_{\rm g}}     \text{,} 
\end{eqnarray}
\noindent
where $r_{\rm g}=2GM_{\rm BH}/c^{2}$ is the Schwarzschild radius, and the $1/(r-r_{\rm g})$ mimics the effects of general relativity (\citealt{paczynsky80}). $M_ {*}$, $R_{\rm eff}$, and $M_{\rm BH}$ are the three major parameters in our model.

In all our simulations, we choose the same initial conditions of hot gas, mimicking the temperature and density profiles derived from X-ray observations \citep{humphrey06}. For the temperature profile, we use the following analytic fit:
\begin{eqnarray}
T(r)=& T_{0} &- (T_{0}-T_{1})e^{-r/(2r_{c})}    \text{,} \label{fit1} 
\end{eqnarray}
where $T_{0}=1.3$ keV, $T_{1}=0.6$ keV, and $r_{c}=8$ kpc. The density profile is then solved assuming hydrostatic equilibrium in the observationally-constrained gravitational potential with $M_ {*}=2.64\times10^{11}M_{\sun}$, $R_{\rm eff}=3.4$ kpc \citep{humphrey06}, and $M_{\rm BH}=4.9\times10^{8}M_{\sun}$ \citep{ferrarese96}. The adopted initial temperature and density profiles are shown as solid lines in Figure \ref{plot1}.

Assuming spherical symmetry, we use the ZEUS-3D hydrodynamic code \citep{stone92} in its one-dimensional mode. The simulations cover a very large radial range from $r_{\text{min}}=10$ pc to $r_{\text{max}}=1$ Mpc. As shown in \citet{guo14}, a small value of $r_{\text{min}}$ and the contributions of the central galaxy and the SMBH to the gravitational potential are very important in correctly predicting the development of cooling flows in central regions. In order to resolve adequately the inner regions, we adopt a logarithmically spaced grid in which $(\Delta r)_{i+1}/(\Delta r)_{i}=(r_{\text{max}}/r_{\text{min}})^{1/N}$, where $N=1000$ is the number of active zones. The smallest grid near the inner boundary has a spatial size of $\sim0.1$ pc. Each simulation was run for about $1.6$ Gyr.

 \section{Results}
\label{section3}

\subsection{The Bimodal Development of Cooling Flows}

   \begin{figure}
   \centering
\plotone {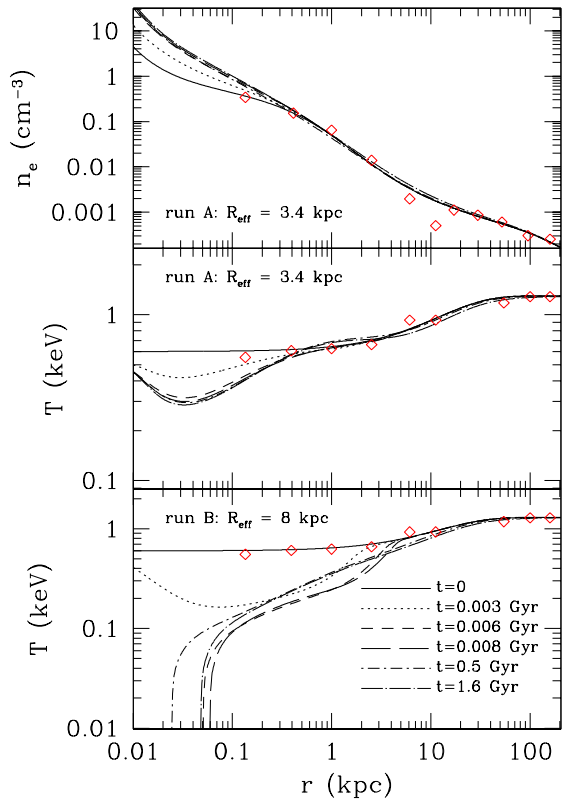} 
\caption{{\it Top and Middle}: Evolution of the gas temperature and electron number density profiles in run A. The diamonds correspond to {\it Chandra} data of NGC 4261 \citep{humphrey06}. The hot gas remains in the hot mode with a cuspy central temperature profile until the end of simulation. {\it Bottom}: Evolution of the gas temperature profile in run B with a larger value of $R_{\rm eff}=8$ kpc. The hot gas develops a central cooling catastrophe very quickly at $t\sim 0.006$ Gyr.}
 \label{plot1}
 \end{figure} 
 
At the beginning of our simulations ($t=0$), the hot gas is static with the same observational temperature and density profiles. Due to radiative cooling, the gas entropy gradually drops and the gas density increases, resulting in a gas inflow toward the center. Even in the simplest spherically-symmetric case with only cooling considered, the development of cooling flows is bimodal \citep{guo14}. In this letter, we performed a large suite of around 90 simulations to explore the parameter space ($M_ {*}$, $R_{\rm eff}$, and $M_{\rm BH}$). Two representative simulations are presented below to showcase the two different fates of cooling flows.

In our fiducial run (denoted as run A), the model parameters are observationally constrained: $M_ {*}=2.64\times10^{11}M_{\sun}$, $R_{\rm eff}=3.4$ kpc \citep{humphrey06}, and $M_{\rm BH}=4.9\times10^{8}M_{\sun}$ \citep{ferrarese96}. The evolution of the hot gas density and temperature profiles are shown in the top and middle panels of Figure \ref{plot1}, indicating that the hot gas reaches a quasi-steady state with a cuspy central temperature profile quickly at $t\sim 0.006$ Gyr, and remains in this hot mode until the end of simulation $t=1.6$ Gyr. Although the hot gas in central regions has a very short cooling time (e.g., $\sim 5$ Myr at $r=0.1$ kpc), the gas still remains in the hot mode quasi-steadily as the compressional heating rate in the spherical inflow becomes comparable to the cooling rate (the inflow timescale is also short), as discussed in detail in \citet{guo14}.
 
As a comparison, the bottom panel of Figure \ref{plot1} shows the temperature evolution of another simulation (denoted as run B) with a larger value of $R_{\rm eff}=8$ kpc (all the other model parameters are the same as run A). The larger $R_{\rm eff}$ results in a less efficient gravitational acceleration ($g(r)\propto M_{*}/(r+a)^{2}$, where $a=R_{\rm eff}/1.8153$) and compressional heating in central regions. As clearly shown in Figure \ref{plot1}, the development of cooling flows in run B is dramatically different from run A. Instead of maintaining a long-term hot mode in run A, the hot gas in run B reaches a central cooling catastrophe very quickly at $t\sim 0.006$ Gyr. Although the cooling catastrophe is expected to be quickly averted by AGN feedback in realistic systems, short epochs of cooling catastrophe may still deposit cold gas to central regions, feeding the growth of the galaxy bulge and the SMBH.

\subsection{The Connection between Star Formation Quenching and Galaxy Structure}
   \begin{figure}
   \centering
\plotone {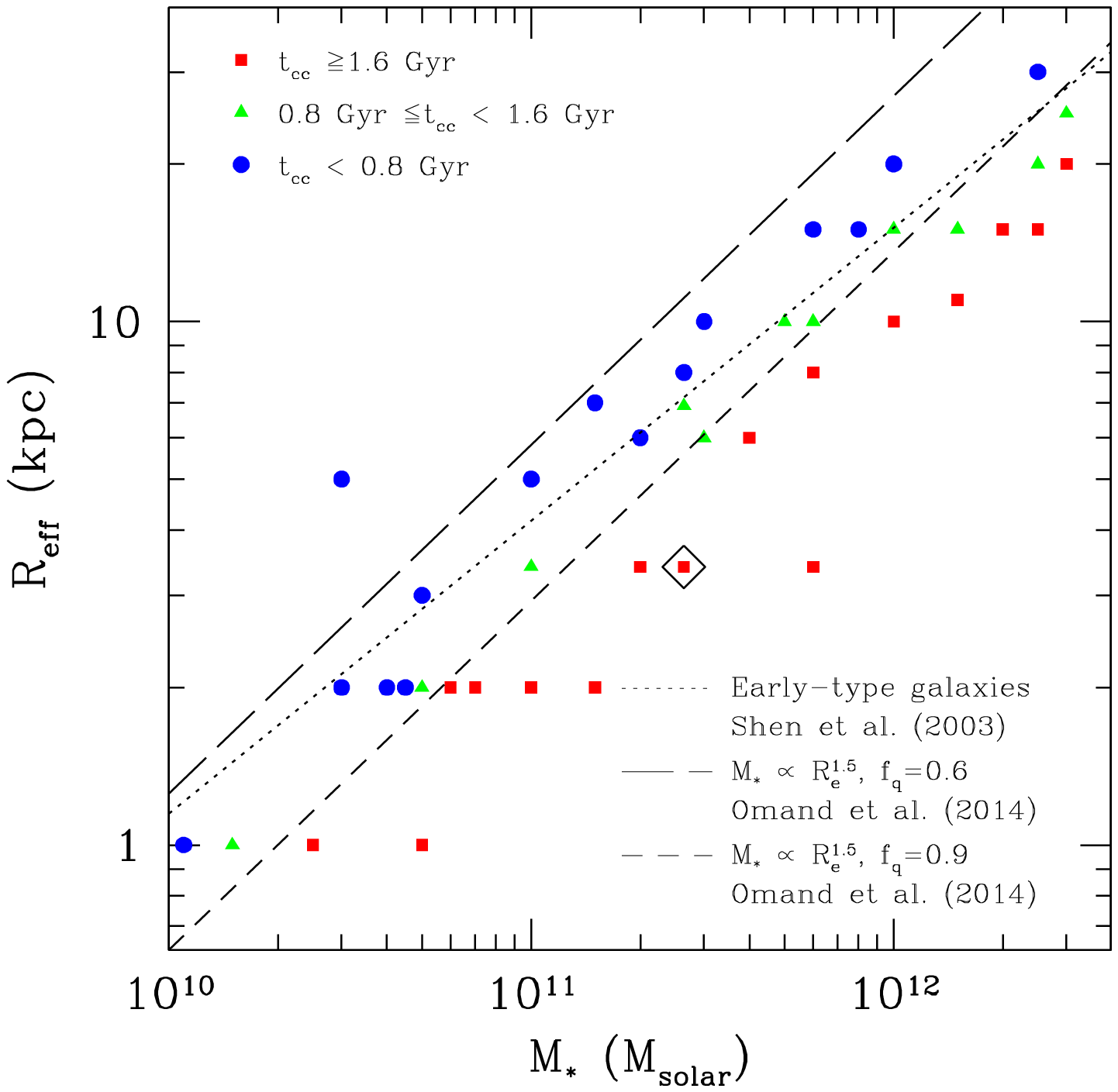} 
\caption{The time when the central cooling catastrophe happens ($t_{\rm cc}$) as a function of $M_{*}$ and $R_{\rm eff}$. The diamond encloses the observational values of $M_{*}$ and $R_{\rm eff}$. The dotted line is the observed $M_{*} - R_{\rm eff}$ relation of SDSS early type galaxies \citep{shen03}. The short and long dashed lines correspond to $M_{*}/R_{\rm eff}^{1.5}=constant$, a structural parameter adopted by \citet{omand14} to discriminate between SF and quiescent galaxies, and the normalization of each line is set by the fraction of quiescent central galaxies $f_{\rm q}$ in \citet{omand14}. }
 \label{plot2}
 \end{figure}

We assume that while quiescent galaxies are maintained in the red sequence (e.g., by radio-mode AGN feedback), the hot halo gas can still form cooling flows intermittently. If the intermittent cooling flows result in a central cooling catastrophe quickly as in run B, cold gas is then deposited onto central regions (within few kpc) and stars form there, enhancing galactic bulges. If a galaxy's bulge is strong, the hot-mode cooling flow can be maintained over a rather long timescale or quasi-steadily due to efficient gravitational heating as in run A. Thus, quiescent galaxies are expected to build up prominent bulges by the central cooling catastrophe in intermittent cooling flows, and galaxy structure (bulge prominence) becomes a good predictor of galaxy quenching. This scenario also naturally leads to a correlation between $M_{*} $ and $R_{\rm eff}$ for quiescent galaxies, corresponding to a central gravitational potential well marginally maintaining the cooling flows in the hot mode quasi-steadily.

\begin{figure}
   \centering
\plotone {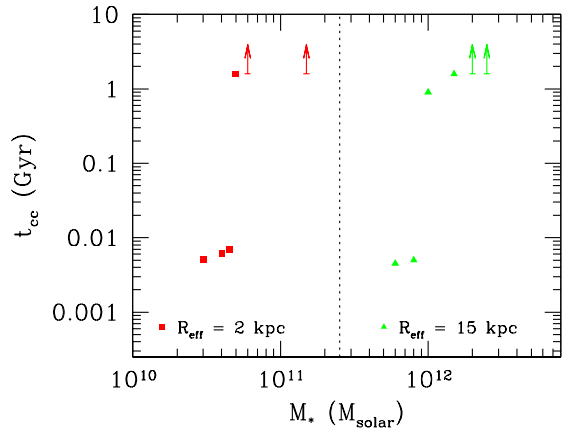} 
\caption{$t_{\rm cc}$ as a function of $M_{*}$ at a fixed value of $R_{\rm eff}$ ($2$ kpc for squares and $15$ kpc for triangles). The transition of the cooling flow fate from cold-mode to hot-mode occurs abruptly with increasing $M_{*}$ (i.e., the depth of the inner gravitational potential well).}
 \label{plot3}
 \end{figure}

To test this scenario, in this subsection we presented a large series of  $\sim 40$ simulations with varying values of $M_{*}$ and/or $R_{\rm eff}$, while keeping the SMBH mass fixed to be the observational value $M_{\rm BH}=4.9\times10^{8}M_{\sun}$. Covering over two orders of magnitude in $M_{*}$, the results are summarized in Figure \ref{plot2}, which shows the time when the central cooling catastrophe happens ($t_{\rm cc}$) as a function of $M_{*}$ and $R_{\rm eff}$. Here each dot represents a specific run, and all simulations are classified into three categories. The red rectangles refer to simulations maintaining the hot cooling flow without producing the central cooling catastrophe within the simulation time (i.e., $t_{\rm cc}\ge 1.6$ Gyr). The green triangles refer to simulations with $0.8$ Gyr $\leq t_{\rm cc}< 1.6$ Gyr, while the blue circles refer to simulations producing the cooling catastrophe quickly with $t_{\rm cc}< 0.8$ Gyr. It should be noted that the transition of the cooling flow fate from the cold mode to hot mode occurs abruptly with increasing the depth of the inner gravitational potential well, as seen in Figure \ref{plot3}.
 
Interestingly, the green triangles, which separate the red rectangles (hot mode) and the blue circles (quick cooling catastrophe), are nicely located along a line with a nearly constant value of $M_{*}/R_{\rm eff}^{1.5}$ (the short dashed line), which is actually the best structural parameter found in \citet{omand14} to distinguish between star forming and quiescent SDSS galaxies. Furthermore, the normalization of the short-dashed line is set by $M_{*}/R_{\rm eff}^{1.5}=10^{10.3}M_{\sun}/\text{kpc}^{1.5}$, corresponding to the quiescent fraction of SDSS central galaxies $f_{\rm q}=90\%$ shown in Figure 12 of \citet{omand14}. The long dashed line with the same slope passes through blue circles, referring to a quiescent fraction of $f_{\rm q}=60\%$. The dotted line, located between these two dashed lines, corresponds to the observed $R_{e}-M_ {*}$ relation of SDSS early-type galaxies \citep{shen03}:
\begin{eqnarray}
R_{\rm eff}= 2.88\times 10^{-6}\left(\frac{M_{*}}{M_{\sun}}\right)^{0.56}     \text{~kpc.}  
\end{eqnarray}
Note that the slope of this relation is close to that of a constant effective stellar surface density $M_{*}/R_{\rm eff}^{2}$, which is the structural parameter suggested by \citet{kauffmann03} and \citet{kauffmann06} to distinguish between star forming and quiescent galaxies.

It is remarkable that our model naturally predicts that $M_{*}/R_{\rm eff}^{1.5}$ is a very good structural parameter discriminating between star forming and quiescent galaxies, confirming observational results of \citet{omand14}. Considering the simplifying assumptions in our model, $M_{*}/R_{\rm eff}$ and $M_{*}/R_{\rm eff}^{2}$ may also be good structural predictors of quenching. The predicted threshold of $M_{*}/R_{\rm eff}^{1.5}=10^{10.3}M_{\sun}/\text{kpc}^{1.5}$ is also consistent with observations of quenched galaxies. The long dashed and dotted lines pass through blue circles, indicating that the transition from star forming to quiescent galaxies starts at lower values of $M_{*}/R_{\rm eff}^{1.5}$. This may suggest that the typical timescale between two successive feedback heating events is less than 0.8 Gyr, the average gas metallicity is less than $0.8Z_{\sun}$ (less efficient cooling), or additional heating mechanisms (e.g. stellar feedback) are playing a role.
 
\subsection{The Role of Supermassive Black Holes}

   \begin{figure}
   \centering
\plotone {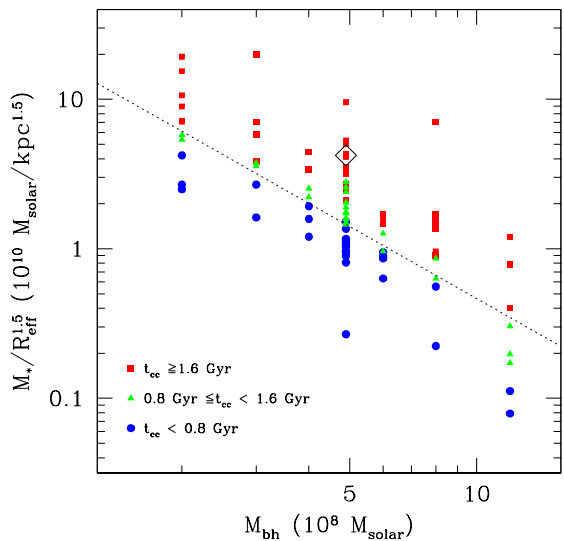} 
\caption{$t_{\rm cc}$ as a function of the SMBH's mass $M_{\rm bh}$ and $M_{*}/R_{\rm eff}^{1.5}$. The diamond encloses the fiducial values adopted from observations. The dotted line passes through green triangles with a slope of $-1.6$.}
 \label{plot4}
 \end{figure} 

The SMBH located at the galactic center usually dominates the gravitational acceleration within the central tens or hundreds pc. Thus its mass may also affect gravitational heating and the fate of cooling flows. We explored this effect with an additional set of around 50 simulations with varying values of $M_{\rm bh}$, $M_{*}$, or $R_{\rm eff}$. Together with the simulations in Section 3.2, the results are presented in Figure \ref{plot4}, which shows $t_{\rm cc}$ as a function of $M_{\rm bh}$ and $M_{*}/R_{\rm eff}^{1.5}$, where the latter represents the impact of galaxy structure investigated in Section 3.2. 

In Figure \ref{plot4}, we show a dotted line roughly passing through green triangles, which separate red rectangles (hot mode) and blue circles (quick cooling catastrophe). The negative slope ($-1.6$) indicates that the SMBH's mass helps maintain cooling flows in the long-term hot mode, and the required minimum significance of galaxy bulge ($M_{*}/R_{\rm eff}^{1.5}$) drops with increasing $M_{\rm bh}$. Due to the bulge buildup through cooling catastrophe, quiescent galaxies are expected to lie along or above the dotted line:
\begin{eqnarray}
\frac{M_{*}/R_{\rm eff}^{1.5}}{10^{10}M_{\sun}/\text{kpc}^{1.5}}= 18.5\left(\frac{M_{\rm bh}}{10^{8}M_{\sun}}\right)^{-1.6}     \text{~.}  
\end{eqnarray}

Our calculations thus predict a general predictor of quenching to be $M_{\rm bh}^{1.6}M_{*}/R_{\rm eff}^{1.5}$. The threshold value is $\sim18.5(10^{8}M_{\sun})^{1.6}(10^{10}M_{\sun})/\text{kpc}^{1.5}$, and may be lower if cooling flows are heated by additional sources (e.g., stellar feedback). 

 \section{Summary and Discussion}
\label{section:conclusion}

In this letter, we propose that intermittent cooling flows may be responsible for the observed connection between SF quenching and galaxy structure. We assume that quenched galaxies are maintained quiescent by a feedback mechanism, which however still allows cooling flows to develop intermittently in the hot halo gas.

We presented a large suite of around 90 hydrodynamic simulations to study how the development of cooling flows depends on $M_{*}$, $R_{\rm eff}$, and $M_{\rm BH}$. Our calculations show that the fate of cooling flows is bimodal (cold vs hot mode), depending significantly on the gravitational potential well in the central few tens of kpc, usually dominated by the central galaxy. If the galaxy's gravity is not strong enough, cooling flows result in a central cooling catastrophe, supplying cold gas and feeding SF to galactic bulges. When the bulge grows to be significant enough, the gravitational potential well becomes deep and, by effectively offsetting radiative cooling, compressional heating maintains cooling flows in a long-term hot mode with cuspy central temperature profiles. 

Our model thus predicts a self-limited growth mechanism for galaxy bulges (and SMBHs) through intermittent cooling flows, and naturally explains the connection between quenching and bulge prominence. In particular, our calculations demonstrate that $M_{*}/R_{\rm eff}^{1.5}$ is a good structural predictor of quenching. We also show that the SMBH's gravity helps maintain cooling flows in the hot mode and predict a general predictor of quenching to be $M_{\rm bh}^{1.6}M_{*}/R_{\rm eff}^{1.5}$. The predicted connection between SF quenching and $M_{\rm bh}$ may be used to test our model.

The importance of the proposed mechanism depends on the bulge growth rate when the central cooling catastrophe happens. The mass inflow rate in intermittent cooling flows depends significantly on the halo gas densities, typically ranging from $\sim 1$ to several hundreds $M_{\sun}/$yr. While galaxy quenching may result from more energetic events (e.g., major mergers or strong AGN activity) which shut off SF on a timescale shorter than the typical cold gas consumption time, intermittent cooling flows may be an important piece of physics in galaxy evolution, and may be more important at high redshifts when gas densities are higher.

Our calculations adopted several simplifications. We used a fixed gas metallicity $Z\sim 0.8Z_{\sun}$. We additionally performed around 20 simulations with lower ($0.6Z_{\sun}$) and higher ($Z_{\sun}$) metallicities, and found that our main result on the connection between the central cooling catastrophe and galaxy structure is robust. Higher metallicity leads to higher cooling rates, requiring a stronger galactic bulge or higher SMBH mass to prevent the cooling catastrophe, and a 25\% variation in metallicity leads to around 40\% ($\sim 0.15 - 0.2$ dex) uncertainty in the threshold correlations in Figures 2 and 4. 
 
We ignore the variation of gas density and temperature profiles across different systems, which may affect the predicted threshold correlations in Figures 2 and 4. Larger systems tend to have higher gas densities, which increase cooling rates and the value of $\alpha$ in the quenching predictor $M_{*}/R_{\rm eff}^{\alpha}$. This effect may be partially counteracted by the trend of larger $M_{\rm bh}$ in larger systems, which enhances gravitational heating from the SMBH. The extreme case is in massive galaxy clusters, where without additional heating, cooling flows are so strong that gravitational heating usually can not compete with cooling \citep{guo14}, indicating that $M_{*}/R_{\rm eff}^{\alpha}$ is not a good quenching predictor for massive central galaxies there, as observationally suggested in Figure 9 of \citet{omand14}. Another limitation of our calculations is the neglect of angular momentum, which may not be significant in the hot halo gas and may be transported outward as the gas flows inward (e.g., by viscosity; \citealt{narayan11}). If important, angular momentum may lead to the formation of extended cold-gas disks when cooling catastrophe develops, resulting in larger values of $R_{\rm eff}$ and $M_{*}$ in quenched galaxies. These limitations should be further investigated in future studies.

\acknowledgements 

FG acknowledges the support by the Zwicky Prize Fellowship of ETH Z\"{u}rich. FG thanks Simon Lilly for insightful discussions and an anonymous referee for many helpful comments.

\label{lastpage}

\end{document}